\newcommand{\ep}{\epsilon}

\newcommand{\om}{\omega}

\newcommand{\si}{\sigma}
\newcommand{\Si}{\Sigma}

\newcommand{\p}{\partial}

\newcommand{\beq}{\begin{equation}}
\newcommand{\eeq}{\end{equation}}
\newcommand{\bna}{\begin{eqnarray}}
\newcommand{\ena}{\end{eqnarray}}
\newcommand{\ba}{\begin{eqnarray*}}
\newcommand{\ea}{\end{eqnarray*}}
\newcommand{\ben}{\begin{enumerate}}
\newcommand{\een}{\end{enumerate}}
\newcommand{\bi}{\begin{itemize}}
\newcommand{\ei}{\end{itemize}}

\documentstyle[preprint,aps,prb,epsf]{revtex}
\begin{document}
\preprint
\widetext

\title{Magnetic Phase Diagram of Hubbard Model in Three Dimensions:\\
       the Second-Order Local Approximation
      }

\author{A.N.\  Tahvildar-Zadeh$^1$ , J.K.\ Freericks $^2$ and M.\ Jarrell
$^1$\\ }
\address{
         $^1$Department of Physics,
University of Cincinnati, Cincinnati, OH 45221\\
         $^2$Department of Physics,
Georgetown University, Washington, DC 20057-0995\\
 	}

\date{\today}

\maketitle
\begin{abstract}

A local, second-order (truncated) approximation is applied
to the Hubbard model in three dimensions.  Lowering the temperature, at
half-filling, the paramagnetic ground state becomes unstable towards the
formation of a
commensurate spin-density-wave (SDW) state (antiferromagnetism)  and
sufficiently far away from half-filling towards the formation of incommensurate
SDW
states.
The incommensurate-ordering wavevector does not deviate much from the
commensurate one, which is in accord with the experimental data for the SDW in
chromium alloys.
\end{abstract}
\renewcommand{\thefootnote}{\alpha{footnote}}
\pacs{74.20.-z, 71.27.+a, 71.38.+i}
\narrowtext
\section{Introduction}

Pure Cr has a spin-density-wave (SDW) ground state with an ordering
wavevector which is incommensurate with the underlying lattice\cite{fawcett}.
Adding electrons
to Cr (by alloying with Mn) makes the magnetic order commensurate with the
lattice and increases the transition temperature, whereas removing electrons
from the system (by alloying with V) drives the magnetic order to a more
incommensurate one, and decreases the transition temperature, eventually to
zero
(Fig. 5).
Penn\cite{penn} found qualitatively the same behavior for the ground state of a
 single-band Hubbard model within a mean-field approximation.  The mean-field
approximation leads to the usual Stoner criterion for the instability of
the paramagnetic state to the  formation of a SDW state.
A first-order perturbation expansion of the  self-energy  in terms of the
interaction parameter leads to the same criterion.
In this contribution we extend this approximation one step further to a
second-order (non-self-consistent)
expansion for the self-energy which includes the lowest-order quantum
fluctuations. But, motivated by the case for large spatial dimensions, we
ignore the nonlocal (site-non-diagonal) elements of the
self-energy\cite{Metzner}.
In fact, inclusion of the lowest-order quantum fluctuations has been shown to
have a dramatic effect on transition temperatures and phase diagrams in the
large-dimensional limit, and has been shown to agree with the quantum Monte
Carlo data (QMC) over a wide range of interaction
strengths\cite{VanDongen,jarrell,freericks}.

The Hubbard model\cite{hubbard} is perhaps the simplest model which can be used
to study the many-body aspects of correlated electrons on a lattice. The
Hubbard Hamiltonian is written in the form
\bna
{\cal H} = -t \sum_{<i,j>,\si} (c_{i,\si}^{\dag} c_{j,\si} + c_{j,\si}^{\dag}
c_{i,\si} ) + U \sum_{i} n_{i,\uparrow} n_{i,\downarrow}
\ena
where $c_{i,\si}^{\dag} (c_{i,\si})$ represents the creation (destruction)
operator of an electron in a Wannier state of spin $\si$ ($\si = \pm {1 \over
2} \, or
\uparrow \downarrow$), on site $i,$ and
$n_{i,\si} = c_{i,\si}^{\dag} c_{i,\si} $ is the electron number operator. The
first term corresponds to the kinetic energy and describes the hopping of
electrons between nearest-neighbor sites on a lattice via an overlap integral
t. This term gives a tight-binding description of the electrons in a periodic
potential forming a single energy band $\ep_{\si}({\bf k}) = -2t\sum_{i=1}^3
\cos k_ia$ for a simple cubic lattice with lattice constant $a$ .
The second term corresponds to the Coulomb repulsion between  electrons. The
long-range Coulomb interaction is assumed to be screened in the solid so that
only the interaction between two  electrons on the same site is retained,
yielding the additional energy of $U$ when the lattice site is doubly occupied.

The model is specified by three parameters: the strength of the electron
interaction $U$ (measured relative to $t$);
the electron density per spin or electron-filling
$n_e={1\over 2N} \sum_{i,\si} <c_{i,\si}^{ \dag} c_{i,\si}>,$
where $N$ is the number of sites in the lattice; and the temperature $T$.

In Section II we introduce the formalism and the approximation that we  use to
form the phase diagram.
In Section III, the details of the numerical calculations are described.
Section IV, presents the results for the second-order approximation and
compares them to the first-order approximation, and the QMC results. A
semiquantitative comparison is also made with the experimental data for Cr.
Conclusions follow in Section V.

\section{Formalism }

For a given $U$ and $n_e$ there may exist more than one type of spin order for
the
ground state, each being stable at a different temperature.  Here we  start
from the paramagnetic state  and find a criterion for
the instability towards the formation of a SDW state. To do this
we couple an external magnetic field to the system and look for singularities
in the response function (magnetic susceptibility) as we change the model
parameters.
The total Hamiltonian of the system in the presence of the magnetic field $h_i$
is
\bna
{\cal H}_h={\cal H} - \sum_i h_i S_i^z.
\ena
where ${\cal H}$ is the Hamiltonian in the absence of the external field
[found in Eq.~(1)] and $i$ is the position of the $i$th electron with spin
$S_i^z=\sum_{\si} \si n_{i\si}$.
The spatial variation of the external field $h_i$ is chosen to probe the
particular expected order for
the spins. For example, if we want to examine the instability towards
antiferromagnetism we choose  $h_i$ to be of the same magnitude everywhere but
of the opposite sign on the two sublattices of the bipartite lattice.

The static response of the system (or the static spin susceptibility) at
temperature $T$ is defined as follows:
\bna
\chi_{ij}&=&2{\p <S_i^z> \over \p h_j} | _{h=0}, \\
&=&2T \sum_{n,\si} \si {\p {\cal G}_{ii}^{n,\si} \over \p h_j} |_{h=0} e^{-i
\om_n 0^-},
\ena
where ${\cal G}_{ij}^{n,\si}=\int d\tau e^{-i\om_n\tau } <T_\tau c_{j,\si}(\tau
) c_{i,\si}^{\dag}(0)> , $ is Green's function at the Matsubara
frequency $\om_n=(2n+1)\pi T$.

Dyson's equation for Green's function (of the Hamiltonian ${\cal H}_h$) is,
\bna
({\cal G}^{n,\si})^{-1}_{ij} = ({{\cal G
}^{0n,\si}})^{-1}_{ij}-\Si^{n,\si}_{ij}+\si h_i\delta_{ij},
\ena
where ${\cal G }^{0n,\si}$ is the noninteracting ($U=0$) Green's function and
$\Si^{n,\si}_{ij}$ is the matrix element of the proper self-energy.
Relation (5) and the derivative of the following identity,
\bna
({\cal G}^{n,\si})_{ij} = ({\cal G}^{n,\si})_{il} ({{\cal
G}^{n,\si}})^{-1}_{lm}
({\cal G}^{n,\si})_{mj},
\ena
are employed to find,
\bna
\chi^n_{ij} = \chi^{0n}_{ij}+2\sum_{k,k',l,l'}\sum_{n',\si,\si'} \si
 {\cal G}^{n,\si}_{ik} {\cal G}^{n,\si}_{li} {\delta \Si^{n,\si}_{kl} \over
\delta {\cal G}^{n',\si'}_{k'l'}} {\p {\cal G}_{k'l'}^{n',\si'} \over \p h_j}\
|_{h=0},
\ena
where $\chi_{ij}=T\sum_n \chi^n_{ij}$ and the bare susceptibility satisfies
$\chi^{0n}_{ij}=-\sum_{\si}{\cal G}^{n,\si}_{ij}{\cal G}^{n,\si}_{ji}.$

For the paramagnetic state, we write ${\cal G}^{n,\si}_{ij}={\cal
G}^{n,-\si}_{ij}$ and ${\delta \Si^{\si} \over \delta {\cal G}^{\si'}} ={\delta
\Si^{-\si} \over \delta {\cal G}^{-\si'}}$, so that,
\bna
\chi^n_{ij} = \chi^{0n}_{ij}+\sum_{k,k',l,l'}\sum_{n'}
 {\cal G}^{n}_{ik} {\cal G}^{n}_{li} \left[ {\delta \Si^{n,\uparrow}_{kl} \over
\delta {\cal G}^{n',\uparrow}_{k'l'}}-{\delta \Si^{n,\downarrow}_{kl} \over
\delta {\cal G}^{n',\uparrow}_{k'l'}} \right] \sum_{\si'} 2\si' {\p {\cal
G}_{k'l'}^{n',\si'} \over \p h_j}\ |_{h=0}.
\ena

We need an explicit form for the self-energy to simplify Eq.~(8)
further. Motivated by the work for large spatial dimensions\cite{VanDongen} we
use a perturbation expansion for the self-energy  up to  second-order in $U/t$.
It was found that the resulting
self-energy is almost local in three dimensions\cite{Schweitzer}, i.e.
$\Si_{ij}\approx\Si_{ii} \delta_{ij},$ so we employ the local approximation
$\Sigma_{ij}=\Sigma_{ii}\delta_{ij}$.   This approximation becomes exact
in large spatial dimensions\cite{Metzner}, and in three dimensions, the effect
of the nonlocal fluctuations on $T_c$ is around 3\% in the weak-coupling
limit\cite{VanDongen}.

Fig. 1 shows the diagrammatic expansion of the local self-energy  through
second order,  which includes the Hartree term and the second-order bubble.
Evaluating the diagrams yields,
\bna
\Si^{m,\si}_{ij} = \left[ UT\sum_n {\cal G}^{n,-\si}_{ii} -U^2 T^2 \sum_{n,n'}
{\cal G}^{n,\si}_{ij} {\cal G}^{n',-\si}_{ij}{\cal G}^{n+n'-m,-\si}_{ji}
\right]
\delta_{ij},
\ena
for the second-order local self energy.
Substituting this approximation in Eq.(8) and Fourier transforming to the
reciprocal lattice, gives a Dyson's-like equation for the susceptibility
\bna
\chi({\bf q}) = \chi^0({\bf q})+ T^2\sum_{n,n'} \chi^{0n}({\bf q})
\Gamma^{n,n'}_{loc} \chi^{n'}({\bf q}) ,
\ena
where $\chi({\bf q})$ is the Fourier transform of $\chi_{ij}$ in the first
Brillouin zone and
\bna
\Gamma_{loc}^{n,n'} &=& U[1-U\chi_{loc}^{pp}(i\om_{n+n'}) ],
\ena
is the irreducible vertex function.
$\chi_{loc}^{pp}$ denotes the local particle-particle susceptibility which is
given by,
\bna
\chi_{loc}^{pp}(i\om_{n}) = {T\over N^2} \sum_{r,{\bf p},{\bf p'}} {\cal
G}^r({\bf p}) {\cal G}^{-r+n}({\bf p'}),
\ena
and $\chi^0({\bf q})$ is the usual bare particle-hole susceptibility,
\bna
\chi^0({\bf q}) = {-T \over N} \sum_{n,{\bf p}} {\cal G}^n({\bf p}) {\cal
G}^n({\bf p}+{\bf q}).
\ena

We expect $\chi({\bf q}_c)$ to diverge at some temperature $T_c$ and
electron-filling $n_{ec}$ when the system is unstable towards the formation of
a SDW at the ordering wavevector ${\bf q}_c$. So, near the transition
temperature, we can neglect $\chi^0({\bf q})$ compared to $\chi({\bf q})$ in
Eq.(10). At low temperatures, the temperature dependence of $\chi_{loc}^{pp}$
is found to be negligible and hence  $\Gamma^{n,n'}$ becomes independent of $n$
and $n'.$ Thus for low
transition temperatures and small $U\chi_{loc}^{pp}$, we find from Eq.(10) and
Eq.(11),

\bna
{1\over U}=\chi^0({\bf q}_c,T_c,n_{ec}) - \chi_{loc}^{pp}(T=0,n_{ec}),
\ena
as the condition for the transition from the paramagnetic phase to an ordered
SDW phase.
This equation is called the modified Stoner criterion\cite{freericks} for the
magnetic instability of the paramagnetic ground state. If the particle-particle
susceptibility is
ignored on the right-hand-side of Eq.~(14), the modified Stoner criterion
becomes the usual Stoner criterion; this  term results
from including the second-order graph in the self-energy expansion.

We must apply one more approximation to calculate the two types of
susceptibilities that appear in modified Stoner criterion . We use bare Green's
functions in Eq.(12) and Eq.(13) to calculate these susceptibilities. In other
words,
we are performing a truncated expansion for the self-energy that is not
self-consistent. This yields
\bna
\chi^0({\bf q})={-T\over (2\pi)^3}\sum_n \int d^3p {1\over i\om_n -\ep({\bf p})
+ \mu} . {1\over i\om_n -\ep({\bf p}+{\bf q}) + \mu},
\ena
and
\bna
\chi^{pp}_{loc}(T=0)=\lim_{T \to 0} {T\over (2\pi)^6}\sum_n \left| \int d^3p
{1\over i\om_n -\ep({\bf p}) + \mu} \right| ^2.
\ena

{\it The lowest-order effect of the quantum fluctuations is remarkably simple:
just reduce the
momentum-dependent particle-hole susceptibility by the local particle-particle
susceptibility before applying the Stoner criterion.}

\section{Numerics}

For each value of $T_c$, ${\bf q}_c$ and $U$, the root of Eq.(14) yields the
critical filling $n_{ec}$ for the SDW order.
We use the particle-hole symmetry of the model to find the phase diagram only
for values of $n_e < 0.5$; the phase diagram is symmetric around $n_e=0.5 .$
For a fixed temperature we can find different roots by changing the ordering
wavevector ${\bf q}_c.$ Since the modified Stoner criterion of Eq.~(14)
is valid only in the nonmagnetic
region of the phase diagram, we accept that value of filling for which an
instability occurs first, as we approach half-filling, i.e., we search
for the ${\bf q}_c$ in the first Brillouin zone which makes the filling minimal
for a fixed temperature.

Calculations of the modified Stoner criterion for finite-size lattices show
that the desired  ${\bf q}_c$ changes only along the edge of the Brillouin zone
as the system is doped away from half-filling, i.e. ${\bf q}_c=(\pi,\pi,q_z)$
for the reduced zone that contains the z-axis. We use this fact to reduce the
triple integral in Eq.(15) to an effectively one-dimensional integral in the
following way: First note that we can write,
\bna
\chi^{0n}({\bf q}) = {-1\over (2\pi)^3} \int&d^3p&\,{1\over 2(i\om_n +\mu) +2t
\cos (p_z)+2t \cos (p_z+q_z)}  \nonumber \\
&.&\left[  {1\over i\om_n -\ep({\bf p}) + \mu} +{1\over i\om_n -\ep({\bf
p}+{\bf q}) + \mu}\right].
\ena
This becomes an integral over the 3D density of states if $q_z=\pi$ (i.e. at
the
zone corner).  Note however that the dependence on $p_x$ and $p_y$ is only
through $-2t \cos (p_x) -2t  \cos (p_y)$ which allows the $p_x$ and $p_y$
integrals
to be replaced by an integral over the 2D density of states, and produces
local Green's functions in 2D:
\bna
\chi^{0n}({\bf q}) = {-1\over 2\pi }&\int&dp_z\, {1\over 2(i\om_n +\mu) +2t
\cos (p_z)+2t \cos (p_z+q_z)}   \nonumber \\
&.&\left[ {\cal G}_{2D}(i\om_n+2t \cos (p_z)+\mu)+{\cal G}_{2D}(i\om_n+2t \cos
(p_z+q_z)+\mu)\right].
\ena
Here ${\cal G}_{2D}(z)$ is the local 2D Green's
function,
\bna
{\cal G}_{2D}(z) = \int d\ep {\rho_{2D}(\ep) \over z-\ep }.
\ena

${\cal G}_{2D}(z)$ is evaluated with a quadrature algorithm which employs a
rational function expansion for large $|z|$, and employs
a 512 point Gaussian integration when $|z|$ is small. So
we can evaluate $\chi^0({\bf q})$ efficiently by numerically performing the
remaining  integration over $p_z$. Note that $\chi^{pp}_{loc}$
is already in the form of a one-dimensional integral over the 3D density of
states.

\section{Results}

Fig. 2 shows the resulting phase boundary for the Hubbard model using the
modified Stoner criterion.
At half-filling the transition is always commensurate with the lattice, i.e.
${\bf q}={\pi \over a}(1,1,1).$ As the system is doped away from half-filling
the transition temperature decreases and a point is reached where the
transition becomes incommensurate with the underlying lattice. Finite-size
calculations show that the ordering wavevector changes only along
the Brillouin zone edge as the transition becomes incommensurate, i.e.
${\bf q}={\pi \over a}(1,1,1-\delta)$ .
The inset in Fig. 2 shows that $\delta$ remains small  and eventually stops
changing as the filling is changed.
Note that the ordering wavevector initially changes very rapidly at the
incommensurate order onset, which is reminiscent of the first-order jump seen
in the Cr
 data\cite{fawcett}.

Fig. 3 shows the result of a finite-size calculation for the original Stoner
criterion which results from a first-order (Hartree) approximation
for the self-energy. We see that the transition temperature is enhanced by
almost a factor of four for each value of $U$ compared to the second-order
approximation results at half-filling. For a given value of $U$ the change in
slope at the onset of incommensuration is much smaller than the results of the
second-order approximation. This change of slope in the second-order
approximation is again similar to what is seen in the Cr data.
Also, as was already shown by Penn\cite{penn}, in the first-order
approximation, for values of $U$ large enough, the ordering wavevector changes
first along the edge of the zone, but then continues to move toward the zone
center as the system is doped far away from half-filling.
This latter behavior has suggested that the system can become ferromagnetic for
low electron filling which is ruled out once the quantum fluctuations are
included.

Fig. 4 shows a comparison of the transition temperature versus $U$ at
half-filling for the first and the second-order approximations as well as the
result of a quantum Monte Carlo (QMC) simulation\cite{scalettar}. We see that
in the case of  the second-order approximation, $T_c$ has negative curvature
for $U/t$ around 10 similar to the QMC result (although the former is not a
valid approximation for large values of $U$), whereas in the first-order
approximation $T_c$ continues to increase with $U$ with a positive curvature.
In fact, since the second-order approximation is exact as $U\rightarrow 0$, it
is clear that the QMC data is overestimating $T_c$ by a factor of four in the
weak-coupling limit.  Such criticism has already been raised by comparing to
other approximation methods\cite{japanese}.

Electronic band structure calculations\cite{fawcett} show that the $d$-electron
concentration for pure Cr is $2.28$ electrons/atom/spin. In order to map
this onto our single-band model, we assume that $1/5$ of this contributes to
our single-band filling, implying $n_e=0.456$ (there are five d-bands in Cr).
The Fermi energy of pure Cr is near the edge of a $d$-band so that the density
of states has a large peak there, thus we assume that doping with Mn adds one
electron to the single  band, and doping with V removes one electron from it.
This maps the rest of the experimental data to the single-band model. We then
fit the value of $U$ in the modified Stoner criterion, to reproduce the
experimental filling for the onset of incommensurate order. As shown in Fig. 5,
the fit is remarkably good for $U=5.5t$. Note that there is only one adjustable
parameter in this fit. The agreement is remarkable since the lattice structure
for Cr is actually bcc and because we made such simple assumptions.
Furthermore, the experimental shape cannot be properly accounted for with a
first-order  approximation for any reasonable value of $U$.

\section{Conclusions}

We found a criterion for the instability of the paramagnetic ground state of
the
3D Hubbard
model towards the formation of an ordered SDW state. This criterion is a simple
modification of the Stoner criterion and is derived by including the
second-order graph in the expansion of the local self-energy. Hence this new
criterion includes the
lowest-order corrections of the old Stoner criterion due to quantum
fluctuations.

We showed that the magnetic phase boundary between paramagnetic and SDW phases
of the model
changes both quantitatively and qualitatively when we include the effect of
quantum fluctuations to the usual mean-field (Hartree) result. The quantum
fluctuations suppress the transition temperatures and exclude the
possibility of the formation of ferromagnetism in the 3D Hubbard model.

We showed that the resulting phase boundary from this modified Stoner criterion
is
consistent with the experimental data for the SDW in dilute chromium alloys
whereas there
are features in the data that cannot be accounted  for by the usual Stoner
criterion with reasonable values of the interaction strength. We were indeed
able to achieve
remarkable agreement with the experimental data, although this model is too
simple to be a realistic one for
Cr. However, since the modified Stoner criterion is a simple and efficient
approximation, we hope it can be applied to more realistic models for Cr
and other transition metals.

\acknowledgments

We would like to acknowledge useful conversations with R. Fishman and P. van
Dongen.
This work was supported at the University of Cincinnati by the National Science
Foundation Grant No. DMR-9406678 and DMR-9357199.
J.~K.~F. acknowledges the Donors of The Petroleum Research Fund, administered
by
the American Chemical Society, for partial support of this research (ACS-PRF\#
29623-GB6).

\begin{figure}[t]
\epsfxsize=6.0in
\epsffile{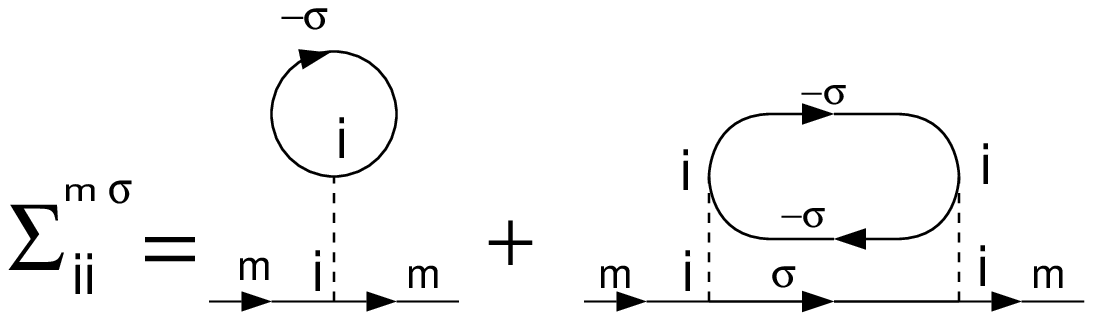}
\caption{The local self-energy $\Si_{ii}$  in the second-order approximation to
the Hubbard model. The Fock term is absent in the Hubbard model. The solid line
represents the undressed $(U=0)$ electron Green's function ${\cal
G}_{ij}^0(i\omega_n)$ and the dotted line represents the intrasite interaction
$U.$ The external legs just show the Matsubara frequency dependence and are not
included in the analytic expressions. }
\end{figure}

\newpage

\begin{figure}[t]
\epsfxsize=6.0in
\epsffile{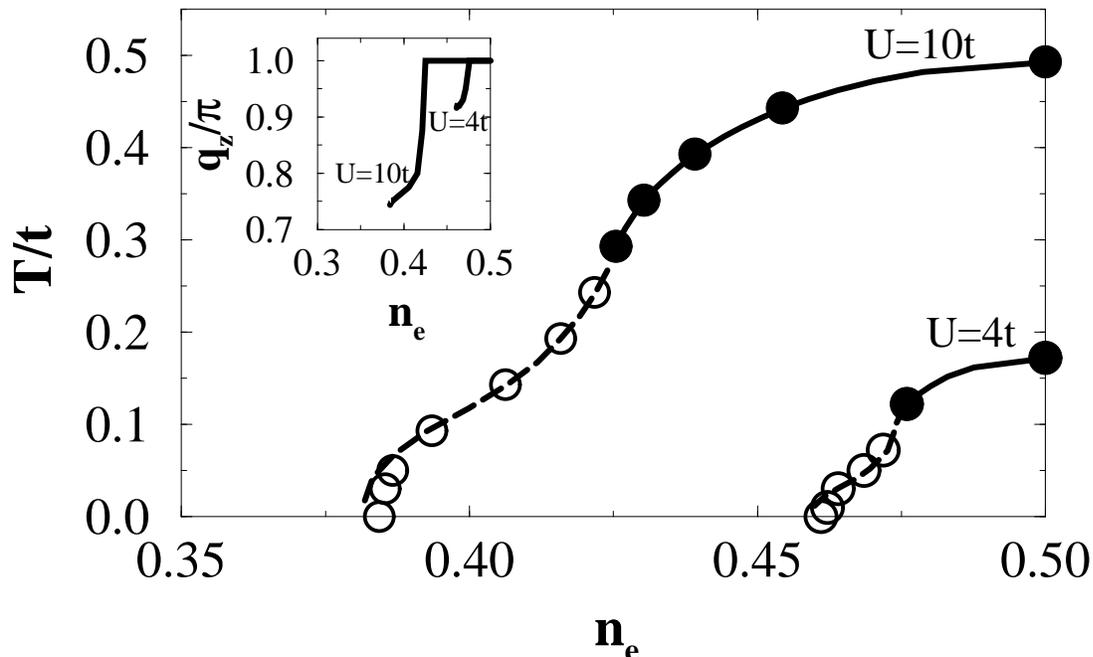}
\caption{Magnetic phase boundaries for the second-order local approximation to
the Hubbard model in three dimensions for two different values of the
interaction strength $U$. The vertical axis shows the ratio of  the temperature
to the hopping constant. The horizontal axis shows the electron filling. The
(dashed) solid lines
denote the (in)commensurate transition from a paramagnetic to a
spin-density-wave ground state in the thermodynamic limit.   The (open) solid
circles denote the corresponding transitions for a finite lattice of 80 unit
cells for $T>0.05t$ and 500 unit cells for lower temperatures.
The inset shows the corresponding z-component of the magnetic ordering
wavevector ${\bf q}$ versus filling. The phase-diagram is symmetric around
$n_e=0.5.$  }
\end{figure}

\newpage

\begin{figure}[t]
\epsfxsize=6.0in
\epsffile{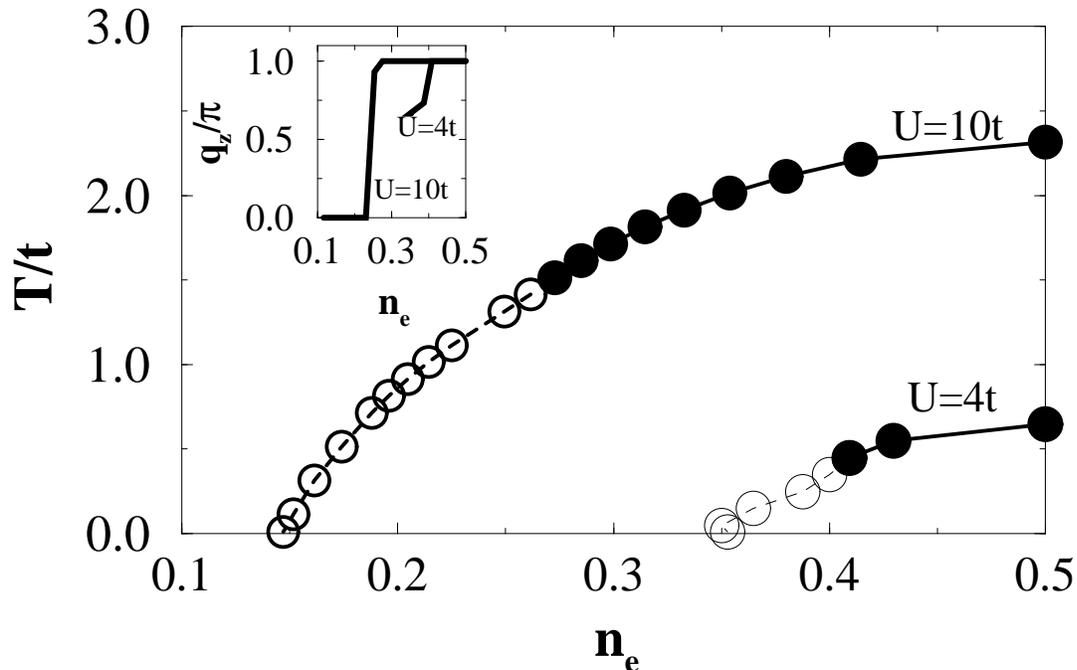}
\caption{Magnetic phase boundaries for the first-order approximation  to the
Hubbard model in three dimensions for two different values of the interaction
strength $U$.
 The vertical axis shows the ratio of  the temperature to the hopping constant.
The horizontal axis shows the electron filling. The (open) solid circles denote
(in) commensurate transition from a paramagnetic to a spin-density-wave ground
state for a finite lattice of 30 unit cells. The lines are a guide to the eye.
The inset shows the corresponding z-component of the magnetic ordering
wavevector ${\bf q}$ versus filling. The phase-diagram is symmetric around
$n_e=0.5.$ }
\end{figure}

\newpage

\begin{figure}[t]
\epsfxsize=6.0in
\epsffile{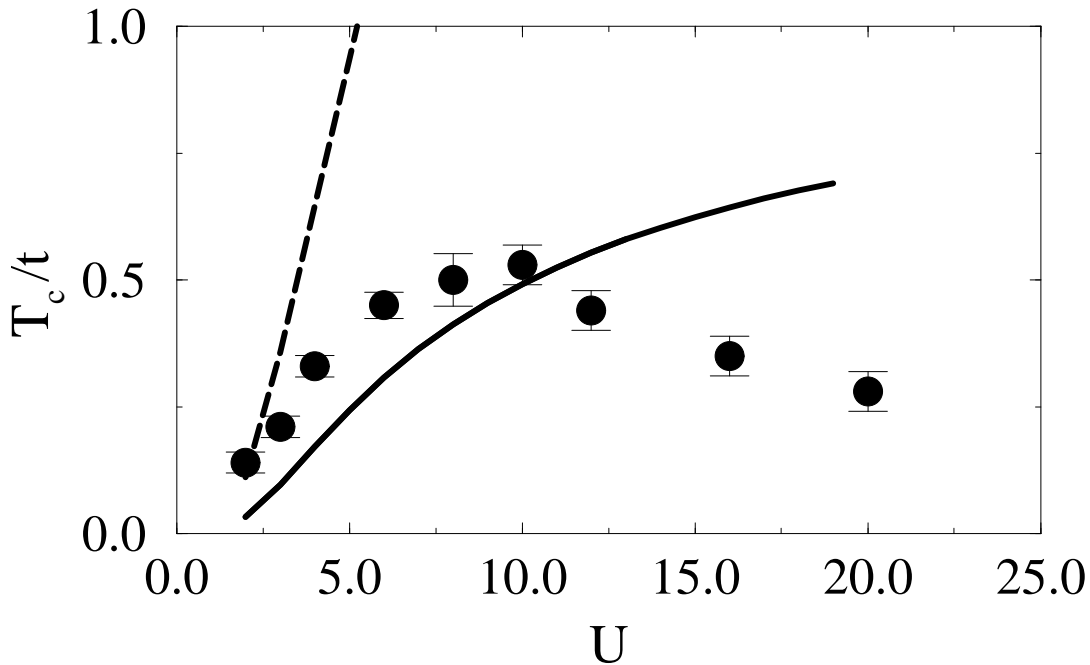}
\caption{Comparison between the results of quantum Monte Carlo simulations$^9$
(circles), the first-order (dashed line)  and the second-order local
approximation (solid line). $T_c/t$ is the ratio of the transition temperature
at half-filling to
the hopping integral. $U$ is the interaction strength. The second-order local
approximation shows the same sign of the curvature as the  QMC result, whereas
the first-order curve shows the wrong sign of the curvature. Also, since the
second-order
approximation has the correct limiting behavior as $U\rightarrow 0$, this shows
that the QMC data is overestimating $T_c$ for small values of $U$.
}
\end{figure}

\newpage

\begin{figure}[t]
\epsfxsize=6.0in
\epsffile{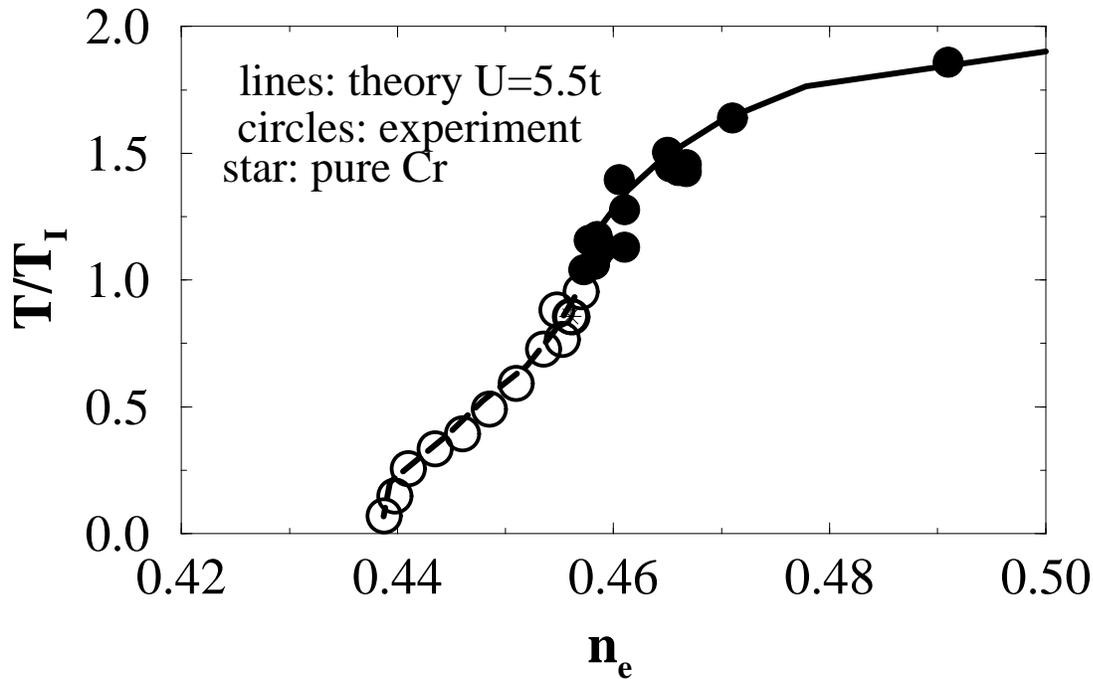}
\caption{Comparison between the experimental data for chromium alloys$^1$ and
the results of the second-order local approximation. The (open) closed circles
denote the experimental data for the (in)commensurate  transitions in Cr
alloys.  The  (dashed) solid lines denote the results of the second-order local
approximation for the (in)commensurate transitions. The experimental data is
best fit to the theoretical calculation when $U=5.5t$.  $T_I$ is the transition
temperature at the paramagnetic-commensurate-incommensurate phase boundary.
$T_I=360 K$ for experimental data and occurs when Cr is doped with $0.3$ atomic
percent of Mn.
}
\end{figure}

\end{document}